\documentclass[reprint,
nofootinbib,
 reprint,
 amsmath,amssymb,
 aps, 
]{revtex4-2}

\usepackage{aas_macros}
\usepackage{graphicx}% Include figure files
\usepackage{dcolumn}% Align table columns on decimal point
\usepackage{bm}% bold math
\usepackage{hyperref}% add hypertext capabilities
\usepackage{mathtools}% for special math symbols
\usepackage{bm}

\makeatletter %% <- make @ usable in command names

\newcommand*\Neginternal[3]{\mathpalette\Neg@{{#1}{#2}{#3}}}
\newcommand*\Neg@[2]{\Neg@@{#1}#2}
\newcommand*\Neg@@[4]{%
  \mathrel{\ooalign{%
    $\m@th#1#4$\cr
    \hidewidth$\m@th#3{#1}\mkern\muexpr#2*2$\hidewidth\cr
  }}%
}
\newcommand*\negslash[1]{\m@th#1\not\mathrel{\phantom{=}}}
\newcommand*\snegslash[1]{\rotatebox[origin=c]{60}{$\m@th#1-$}}
\newcommand*\ssnegslash[1]{\rotatebox[origin=c]{60}{$\m@th#1{\dabar@}\mkern-7mu{\dabar@}$}}
\newcommand*\sssnegslash[1]{\rotatebox[origin=c]{60}{$\m@th#1\dabar@$}}
\makeatother  %% <- revert @

\begin{document}

\preprint{APS/123-QED}

\title{Invariants in Polarimetric Interferometry: a non-Abelian Gauge Theory}

\author{Joseph Samuel}
\email{sam@rri.res.in}
\affiliation{International Centre for Theoretical Sciences, Bengaluru 560089, Karnataka, India} 
\affiliation{Raman Research Institute, Bengaluru 560080, Karnataka, India}%

\author{Rajaram Nityananda}
\email{rajaram.nityananda@gmail.com}
\affiliation{Azim Premji University, Bikkanahalli Main Road, Sarjapura, Bengaluru 562125, Karnataka, India}%
 
\author{Nithyanandan Thyagarajan}%
\email{Nithyanandan.Thyagarajan@csiro.au}
\homepage{https://tnithyanandan.wordpress.com/}
\affiliation{CSIRO, Space \& Astronomy, P. O. Box 1130, Bentley, WA 6102, Australia}%
\affiliation{National Radio Astronomy Observatory, 1003 Lopezville Rd, Socorro, NM 87801, USA}%

\date{\today}% It is always \today, today,

\begin{abstract}
The discovery of magnetic fields close to the M87 black hole using Very Long Baseline Interferometry (VLBI) by the Event Horizon Telescope collaboration utilized the novel concept of ``closure traces'', that are immune to element-based aberrations. We take a fundamentally new approach to this promising tool of polarimetric VLBI, 
using ideas from the geometric phase and gauge theories.
The multiplicative distortion of polarized signals at the individual elements are represented as gauge transformations 
by general $2\times 2$ complex matrices, so the closure traces now appear as gauge-invariant quantities. We apply this formalism to polarimetric interferometry and generalize it to any number of interferometer elements. Our approach goes beyond existing studies in the following respects: (1) we use triangular combinations of correlations as basic building blocks of invariants, (2) we use well-known symmetry properties of the Lorentz group to transparently identify a complete and independent set of invariants, and (3) we do not need auto-correlations, which are susceptible to large systematic biases, and therefore unreliable. This set contains all the information, immune to corruption, available in the interferometer measurements, thus providing important robust constraints for interferometric studies. 
\end{abstract}

%Use showkeys class option if keyword
\maketitle

%\tableofcontents

%\section{Background}\label{sec:intro}

The measurement of coherence of fields is an important concept with applications in many disciplines of physics. Radio astronomers measure
coherence by correlating radio signals received by an interferometer array and characterize the morphology of radio emission received from the sky \cite{TMS2017}. The signals received at each telescope are usually corrupted due to propagation effects and local imperfections in the array receiver elements. Finding interferometric invariants, quantities that are immune to this corruption, 
and so accurately reflect the true structure of the sky emission, is of significant value. This is the principal focus of this paper. 

The subject of closure phases and closure amplitudes, two popular interferometric invariants in astronomy, has a long history \cite{jen58,Twiss+1960}. Concepts analogous to closure invariants (interferometric invariants defined on closed loops) occur in other areas like speckle interferomentry \cite{Weigelt+1983}, crystallography \cite{Hauptman1991} and quantum mechanics \cite{Bargmann1964,Berry1984}. Our approach, as in the recent work \cite{Thyagarajan+2020c}, is inspired by the geometric phase \cite{Berry1984,Berry1987b} of quantum physics and classical optics \cite{Samuel+1988,Bhandari+1988}. A common thread that ties all these diverse problems together is gauge theory. We adapt the general framework to the  problem of closure invariants in astronomy.

Closure invariants in co-polar correlations, in which signals of the same polarization are correlated between all pairs of telescopes, are well studied in both theory \cite{Lannes1991,TMS2017} and practice \cite{Readhead+1978,Pearson+1984}. In a recent significant extension to polarimetric measurements, ``closure traces'' were introduced and explored in detail for a system of four array elements \cite{Broderick+2020}. These closure traces were used in studying the magnetic properties near the event horizon of the supermassive black hole at the center of the galaxy M87 \cite{eht21-7-abridged}. Generalizing and extending 
this work \cite{Broderick+2020} on polarimetric interferometry and closure invariants, our paper applies the gauge theory framework to the 
group of nonsingular 
$2\times 2$ matrices (known to mathematicians as $\textrm{GL}(2,\mathbb{C})$) and Lorentz groups. We go beyond earlier work by transparently determining the complete and independent set of invariants from an $N$-element interferometer array. We show that auto-correlation measurements, which are highly susceptible to systematic errors\footnote{In radio astronomy, auto-correlations are problematic because of their susceptibility to systematic effects caused by emission unrelated to the object being imaged (\textit{e.g.}, atmospheric emission and radio frequency interference) and the gain errors affecting these strong unwanted signals.}, are not necessary. If available, they are easily included in the scheme. This should be useful in future mapping of polarized sources, especially using long baseline interferometry, where calibration is challenging. 

{\it Notation}.---We use indices $a,b=0,\ldots,N-1$ to label the array elements, and indices $p,q=1,2$ to label the two polarizations. $2\times2$ and $4\times 4$ matrices are shown in uppercase boldface. 2-element vectors and 4-vectors are written using lowercase boldface in regular and italicized fonts, respectively. For 4-vectors and $4\times 4$ matrices written in component form, we use relativity conventions, like the Einstein summation convention for repeated Greek indices, which range over $0,1,2,3$.

%\section{Context: Polarimetric Interferometry}\label{sec:context}
\textit{Polarimetric Interferometry}.---Consider an interferometer array with $N$ elements. The $p$-th component of polarization (in some orthonormal basis) of the electric field incident on element, $a$, is denoted by $e_a^p$,  represented in amplitude and phase by a complex 2-element column vector, $\mathbf{e}_a$. The true correlation matrix between the array elements is obtained via pairwise cross-multiplication and averaging, $\mathbf{S}_{ab}\coloneqq \left\langle \mathbf{e}_a\mathbf{e}_b^\dagger \right\rangle$, where, $\dagger$ denotes a conjugate-transpose operation, and $\langle\cdot\rangle$ denotes the average.

Due to the propagation medium and the non-ideal measurement process, the incoming amplitudes are corrupted by an element-based linear transformation -- a general $2\times 2$ complex matrix, $\mathbf{G}_a$. The off-diagonal entries of $\mathbf{G}_a$ represent leakage between the two polarized receiver channels at array element $a$. The measured amplitudes, $\mathbf{v}_a$, are related to the true amplitudes, $\mathbf{e}_a$, by $\mathbf{v}_a=\mathbf{G}_a \mathbf{e}_a$.

The corrupted correlation matrix for a pair of elements, $(a,b)$, is constructed from the measurements as
\begin{align}
  \mathbf{C}_{ab} &= \left\langle \mathbf{v}_a\mathbf{v}_b^\dagger \right\rangle = \mathbf{G}_a \left\langle \mathbf{e}_a\mathbf{e}_b^\dagger \right\rangle \mathbf{G}_b^\dagger = \mathbf{G}_a \mathbf{S}_{ab} \mathbf{G}_b^\dagger \, . \label{eqn:corrupted-corr}
\end{align}
$\mathbf{G}_a$, $\mathbf{C}_{ab}$, and $\mathbf{S}_{ab}$ are $2\times 2$ matrices. Generically, their eigenvalues are non-zero and we will assume this to be true. These matrices are then invertible and so belong to the general linear group, $\textrm{GL}(2,\mathbb{C})$. Evidently, $\mathbf{C}_{ab}^\dagger=\mathbf{C}_{ba}$.
The auto-correlations, $\mathbf{A}_{aa}\coloneqq \mathbf{C}_{aa}$, are Hermitian and positive definite (strictly positive eigenvalues). We use a special symbol to distinguish them from general cross-correlations, $\mathbf{C}_{ab}$. 

Our objective is to construct quantities which are immune to the corruptions and actually reflect the true coherence properties of the source of radio emission, 
rather than local conditions (represented by $\mathbf{G}_a$) at the interferometer elements. This problem is mathematically related to the ``gauge'' theories of fundamental interactions, such as electromagnetism (EM) or the nuclear force. We regard multiplication of the true signal by the local gains, $\mathbf{G}_a$, as gauge transformations. We wish to eliminate these spurious effects introduced by $\mathbf{G}_a$ and identify a maximal, independent set of gauge-invariant quantities (independent of $\mathbf{G}_a$), which we will call ``closure invariants''. The gains are elements of $\textrm{GL}(2,\mathbb{C})$, which is not unitary, resulting in features not seen in the usual gauge theories of particle physics. The case of co-polar observations is described in detail in \cite{TMS2017,Lannes1991}, where the gains are just nonzero complex numbers, and the gauge group is $\textrm{GL}(1,\mathbb{C})$, which is Abelian --- a commutative group. 
In a companion paper \cite[Paper~I;][]{copolar-invariants}, we have introduced these ideas in the simpler context of $\textrm{GL}(1,\mathbb{C})$, which serves as a stepping stone to the polarized theory.
The current paper deals with full polarimetric measurements governed by the gauge group $\textrm{GL}(2,\mathbb{C})$. An important difference between 
the two cases is that matrices $\textrm{GL}(2,\mathbb{C})$ group do not, in general, commute. This introduces a level of complexity not seen in the co-polar (Abelian) case \cite{copolar-invariants}.

%\section{Accounting of Invariants}\label{sec:numcount}
\textit{Counting Invariants}.---How many independent closure invariants can we expect to find? The number of true cross-correlations, $\mathbf{S}_{ab}$, in the measured cross-correlations, $\mathbf{C}_{ab}$, is equal to the number of array element pairs, $N(N-1)/2$, which are assumed to be non-redundant in this paper. Each $\mathbf{C}_{ab}$ gives us one complex $2\times2$ matrix, with $8$ real parameters. If auto-correlations are also measured, we would have to add $4 n_\textrm{A}$ to this count, since $\mathbf{A}_{aa}$ is Hermitian and has $4$ real parameters, and $n_\textrm{A}$ is the number of true auto-correlations in the measurements. When the object occupies a small fraction of the field of view, all elements would measure essentially the same auto-correlation (with an inherent assumption of spatial stationarity).
Therefore, $\mathbf{S}_{aa}=\mathbf{S}_{bb}$ for all $a,b$, and we can set $n_\textrm{A}$ to zero or one without loss of generality. Thus, if $\mathcal{M}$ is the space of all measured correlations, its dimension is the total number of real parameters in the measured correlations, $\mathrm{dim}(\mathcal{M})=8 N(N-1)/2 + 4\,n_\mathrm{A}$. 
 
\begin{figure}
\begin{center}
\includegraphics[width=1.0\linewidth]{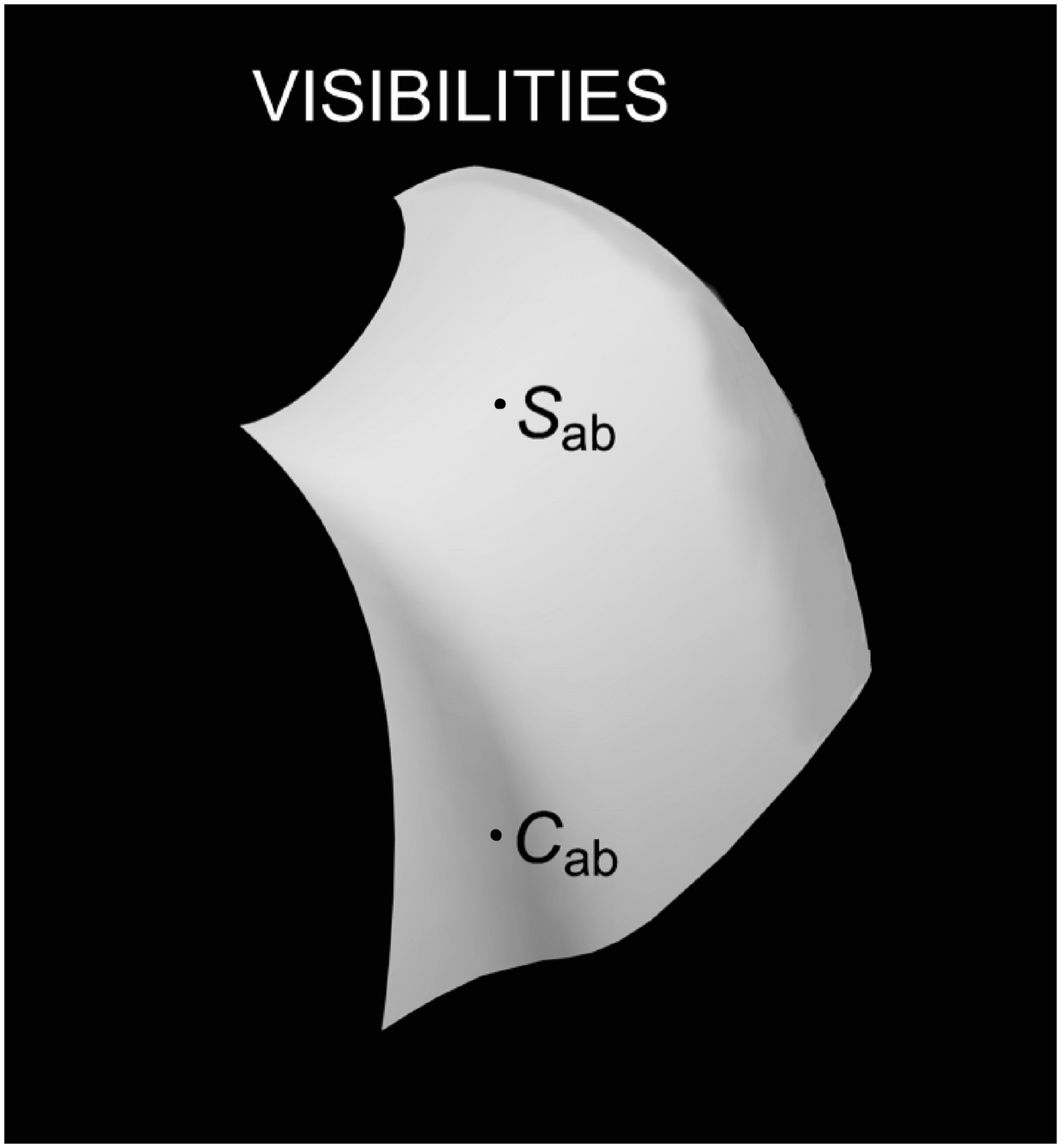}
\caption{A schematic representation of the space of correlations using a dark three-dimensional region. If the true correlation due to the source structure is at the point marked $S_{ab}$, the measured correlation, $C_{ab}$, could lie anywhere on the gray two dimensional surface because of gain distortions. For $N>3, n_\textrm{A}=0$, the correlation space and the surface have real dimensions $4 N(N-1)$ and $8N-1$, respectively. The figure shows these spaces as three- and two-dimensional, respectively, for illustrative purposes.
\label{fig:lattice}}
\end{center}
\end{figure}

Each of the $N$ gain matrices, $\mathbf{G}_a$, is described by 4 complex numbers (or, 8 real numbers), resulting in $8N$ real parameters in total. 
The gains $\mathbf{G}_a$ from the $N$ array elements
corrupt the signals according to Eq.~(\ref{eqn:corrupted-corr}) and the 
$\mathbf{C}_{ab}$ could lie anywhere on a surface of points containing $\mathbf{S}_{ab}$ (see Fig.~\ref{fig:lattice}). Any function of the measurements which is constant on each surface is immune to corruption by the gains. These are the interferometric invariants we seek. The number of independent interferometric invariants is the dimension of $\mathcal{M}$ minus the dimension of the surface.

The dimension of the surface can at most be the dimension of the 
gains $\mathbf{G}_a$, which is $8N$.
It could be smaller if there are gain variations which do not corrupt the measured correlations\footnote{In gauge theory, these correspond to a subgroup (called the stabilizer), which acts trivially (i.e., does not affect the measured correlations).}. We refer to these as Non-Corrupting Gains, NCGs for short. As a result, we would expect
\begin{align}
  N_\mathcal{I} &= 4N^2-12N+4n_\textrm{A}+s \label{eqn:count}
\end{align} 
independent invariants, where $s$ is the number of parameters in 
NCGs (dimension of the stabilizer subgroup). One well-known example of an NCG is a constant phase shift applied to all the signals on all the interferometer elements, which would not affect $\mathbf{C}_{ab}$. Such a uniform phase shift applied to all elements belongs to the $\textrm{U}(1)$ subgroup, which forms the entire stabilizer group. This is the only NCG ($s=1$) in all cases except $N=3,  n_\textrm{A}=0$ (see below).  
For $N=4$ this yields $17$ and $21$ invariants, for $n_\textrm{A}=0$ and $n_\textrm{A}=1$, respectively. We note that our count differs from that of \cite{Broderick+2020}, where NCGs were not considered and the dimension of the surface was assumed to be $8N$ (from $4N$ complex gain parameters).

%\section{Gauge Theory}\label{sec:framework}
\textit{Gauge Theory on a Graph}.---Given the measured correlations, we wish to find a complete and independent set of invariants. In a standard gauge theory, this is done by fixing a base point and considering parallel transport (holonomy) along closed loops which start and end at the base point. 
Such parallel transport measures curvature and is called a  Wilson loop.
Wilson loops generate all the gauge invariant quantities in a gauge theory and are the key to constructing closure invariants.
We fix a base vertex at one of the array elements labeled $0$. We regard each array element, $a$, as a vertex in a graph with $N$ vertices. Each element pair, $(a,b)$, is a directed edge or a ``link'' in the graph, carrying the variable, $\mathbf{C}_{ab}$, a $2\times2$ complex matrix from the $\textrm{GL}(2,\mathbb{C})$ group. This link variable is called a ``connection'' and defines parallel transport from $a$ to $b$. The gains, $\mathbf{G}_a$, are local (\textit{i.e.}, element-based) gauge transformations by the $\textrm{GL}(2,\mathbb{C})$ group. 

In standard gauge theories, such as $\mathrm{SU}(2)$ for example, the gauge group describing $\mathbf{G}_a$ is unitary and Hermitean conjugation in Eq.~(\ref{eqn:corrupted-corr}) is the same as matrix inversion. Since our gauge group is not unitary, we have to proceed differently. So, we define a hat operator on invertible matrices, $\widehat{\mathbf{P}}=(\mathbf{P}^\dagger)^{-1}$ (its relevant properties are listed in appendix~\ref{sec:hatting}). This convenient notation lets us adapt methods employed by \cite{Broderick+2020}. We first define {\it covariants}, $\bm{\mathcal C}_{\Gamma_0}$, as products of an even number of $\mathbf{C}_{ab}$ matrices around a closed loop, $\Gamma_0$, pinned at vertex $0$, with the even terms `hatted'. For example, for $\Gamma_0=0abc0$, $\bm{\mathcal C}_{\Gamma_0}= \mathbf{C}_{0a}\widehat{\mathbf{C}}_{ab}\mathbf{C}_{bc}\widehat{\mathbf{C}}_{c0}$. Under gauge transformations, covariants transform as
\begin{align}
  \bm{\mathcal{C}} &\xmapsto{\mathcal{G}} \mathbf{G}_0 \bm{\mathcal{C}} \mathbf{G}_0^{-1} \, . \label{eqn:covariant-transformation}
\end{align}
Note that by defining covariants pinned at vertex $0$, we have considerably reduced the number of gains contributing to corruptions. 
All that remains is a single gain $\mathbf{G}_0$ at the base vertex 0. Closure invariants can be found by taking traces, $\mathcal{I} = \textrm{tr}(\bm{\mathcal{C}})$, of covariants as in \cite{Broderick+2020} or indeed in gauge theories. These invariants will be unchanged under any gauge transformation with $\mathbf{G}_a$. However, the need for an even number of links  to construct covariants would seem to force us towards rectangles. The problem now is that it is far from clear which set will give us a complete and independent set of  invariants.

Our solution to this problem is to define and use quantities called `advariants', which we introduced in \cite{copolar-invariants}. They are constructed by multiplying an \textit{odd} number of correlations, $\mathbf{C}_{ab}$, around a closed loop, $\Gamma_0$, pinned at vertex $0$ with even terms hatted. For example, for $\Gamma_0=0abcd0$, $\bm{{\mathcal A}}_{\Gamma_0}= \mathbf{C}_{0a}\widehat{\mathbf{C}}_{ab}\mathbf{C}_{bc}\widehat{\mathbf{C}}_{cd}\mathbf{C}_{d0}$. Under gauge transformations, advariants transform as
\begin{align}
  \bm{\mathcal{A}} &\xmapsto{\mathcal{G}} \mathbf{G}_0 \bm{\mathcal{A}} \mathbf{G}_0^\dagger \,, \label{eqn:advariant-transformation}
\end{align}
One special case of an advariant is the auto-correlation, $\bm{\mathcal{A}}_0 \coloneqq \mathbf{A}_{00}$, which obeys Eq.~(\ref{eqn:advariant-transformation}). The next non-trivial example is a 3-vertex advariant defined on an elementary triangle $\Delta_{(0ab)}\equiv (0,a,b)$, $\bm{\mathcal{A}}_{ab} \coloneqq \mathbf{C}_{0a}\widehat{\mathbf{C}}_{ab}\mathbf{C}_{b0}$.
Note that $\bm{\mathcal{A}}_{ab}^\dagger$ is also an advariant. Advariants are the building blocks for covariants. In fact, any  covariant 
can be obtained by multiplying an \textit{even} number of such elementary advariants with even terms hatted as described earlier. For example, a 4-vertex covariant on a closed loop, $\Gamma_0=0abc0$, can be written as
$\bm{\mathcal{C}}_{\Gamma_0} = \bm{\mathcal{A}}_{ab}\widehat{\bm{\mathcal{A}}}_{bc}$.
Note that all these covariants transform in accordance with Eq.~(\ref{eqn:covariant-transformation}) for a closed loop pinned at vertex 0. The traces of these covariants are the closure invariants \cite{Broderick+2020}. 
The key advance in this paper is that we use 
elementary triangular advariants to arrive at
a complete and independent set of closure invariants, 
using general arguments based on Lorentz and scaling symmetries.
An advariant, $\bm{\mathcal{A}}$, can be expanded in terms of the identity matrix, $\boldsymbol{\sigma}_0 \coloneqq \mathbf{I}$, and Pauli matrices, $\boldsymbol{\sigma}_m, m=1,2,3$ as
\begin{align}
  \bm{\mathcal{A}} &= z^\mu \boldsymbol{\sigma}_\mu \, , \label{eqn:advariant-expansion}
\end{align}
where, $z^\mu$ are complex  coefficients. 
The $z^\mu$ can be thought of as similar to complex Stokes parameters $\{I,Q,U,V\}$. 
In the case of the autocorrelation advariant, which is Hermitean, 
these are real and {\it are} the ordinary Stokes parameters of polarization
optics \cite{Britton2000}.

Under gauge transformations of $\bm{\mathcal A}$, the action of the gauge group, 
$\mathbf{G}_0\in \textrm{GL}(2,\mathbb{C})$, on advariants can be split into two parts.
First, the subgroup, $\mathbf{G}_0=\lambda\mathbf{I}$, acts on $z^\mu$ by scaling,
\begin{align}
  z^\mu &\mapsto \left|\lambda\right|^2 z^\mu \, . \label{eqn:scaling}
\end{align}
Second, the subgroup $\textrm{SL}(2,\mathbb{C})$  of matrices with unit determinant 
preserves $\textrm{det}(\bm{\mathcal{A}})= I^2-Q^2-U^2-V^2= z^\mu \eta_{\mu\nu} z^\nu $, which shows that it 
acts on $z^\mu$ by a Lorentz transformation,
\begin{align}
  z^\mu &\mapsto \Lambda^\mu{}_\nu \, z^\nu \,, \label{eqn:lorentz-transform}
\end{align}
where, $\boldsymbol{\Lambda}\equiv \Lambda^\mu{}_\nu$ is a real $4\times 4$ matrix representing a Lorentz transformation.  
Note that the real and imaginary parts of ${\mathbf{z}}$ separately transform as 4-vectors. Closure invariants must remain unchanged under transformations by both these subgroups.

{\it Complete and independent set.}---
The triangular advariants (pinned at vertex 0) can be used to build covariants for every closed loop with even links in the graph. Each triangular advariant, $\bm{\mathcal{A}}_{ab}$, can be expanded as $\bm{\mathcal{A}}_{ab}=z_{ab}^\mu \boldsymbol{\sigma}_\mu$ as in Eq.~(\ref{eqn:advariant-expansion}), where, $z_{ab}^\mu$ are complex  $4-$vectors.

We first consider only the Lorentz transformations in Eq.~(\ref{eqn:lorentz-transform}). Each triangle, $\Delta_{(0ab)}$, gives us a complex 4-vector, $z^\mu_{ab}$, which can be decomposed into two real 4-vectors. The number of triangles with one vertex at $0$ is $N_\Delta=\binom{N-1}{2}$. In addition, for $n_\textrm{A}=1$, there will be one more Hermitian advariant, $\bm{\mathcal{A}}_0$, which gives us one more real 4-vector.
Thus, we have a set of $M=2N_\Delta+n_{\textrm{A}}$ real, independent, 4-vectors, $\boldsymbol{y}_m, m=1,2,\ldots M$.
Invariants of the Lorentz subgroup (Lorentz invariants), can only be functions of Minkowskian inner products among these vectors.

We construct an independent set of closure invariants (for $N>3$) as follows: choose any four real 4-vectors to be the basis set, say $\boldsymbol{h}_k={\bf y}_k, k=1,2,3,4$. 
Between these four independent basis vectors, there are $10$ Minkowskian inner products, $\widetilde{\mathcal{I}}_{k \ell}=\boldsymbol{h}_k\cdot \boldsymbol{h}_\ell$,
with $k,\ell=1,2,3,4$ and $\ell\ge k$. Further, the remaining $(M-4)$ 4-vectors give us $4(M-4)$ Minkowski inner products with the four basis vectors, $\widetilde{\mathcal{I}}_{kn}=\boldsymbol{h}_k\cdot \mathbf{y}_n$
for $n=5,\ldots M$. Together, we have $10 + 4(M-4)  = 4N^2 - 12N + 2 + 4\,n_\textrm{A}$ real Minkowski inner products. There is no need to take any more inner products since a 4-vector is completely specified by its components in a basis.  The $M$ triangular advariants are independent and all covariants can be constructed from these triangular advariants.  This ensures that these Minkowskian inner products are both complete and independent.

The inner products (Lorentz invariants) are still not the invariants we seek since they acquire a scale factor from the element-based gains [Eq.~(\ref{eqn:scaling})]. We can obtain true invariants by  
forming ratios of the Lorentz invariants $\{\widetilde{\mathcal{I}}_{k\ell}, \widetilde{\mathcal{I}}_{kn}\}$, 
for example, dividing the set by any one of them, say $\widetilde{\mathcal{I}}_{11}$. 
Thus, the number of closure invariants will be $4N^2 - 12N + 1 + 4\,n_\textrm{A}$, that is, one less than the number of independent inner products. Comparing with Eq.~(\ref{eqn:count}), we find that $s=1$ (one NCG corresponding to a uniform phase across all elements) for $N\ge 4$.

The algorithmic summary for finding the closure invariants from the measured visibilities is: 1)~Choose a base element $0$ and construct the
$N_\Delta=\binom{N-1}{2}$ complex triangular advariants. 2)~Use these to contruct $2N_\Delta$ real 4-vectors. 3)~If $n_\textrm{A}=1$, include one more real 4-vector
to get $M=2N_{\Delta}+n_\textrm{A}$ 4-vectors.
4)~Pick four of these 4-vectors as a basis and compute the 10 Minkowski dot products between them. 5)~Compute the Minkowski dot products between the remaining 
$(M-4)$ 4-vectors and the basis vectors. 6)~Divide the $10+4(M-4)$ Minkowski dot products by any one of them to find all the closure invariants. Finding 
a complete and independent set of closure invariants in polarized interferometry is the main new result of this paper.

The case $N=3$ is special. For $n_\textrm{A} =1$, we get 5 invariants as expected (from $4N^2-12N+1+4\,n_\textrm{A}$). For $n_\textrm{A}=0$, we may likewise expect to get $1$ invariant. Instead, we get $2$ invariants: 
the three inner products of the two 4-vectors associated with $\bm{\mathcal{A}}_{12}$ give us two invariant ratios. 
The ``extra'' invariant appears because there
is an extra one parameter family of NCGs. These are Lorentz transformations
in the plane orthogonal to the two $4$-vectors. 
We identified a uniform phase shift on all elements as an example of an NCG. Are there others? 
The work of this paper and comparison with Eq.~(\ref{eqn:count}) clearly rules this out in all cases but one ($N=3, n_\textrm{A}=0$). 
While this special case is discussed in appendix~\ref{sec:N=3}, a simple proof of this general statement can be given: 
any Lorentz transformation that preserves all the visibilities also preserves the advariants and the associated $4$-vectors. 
The only Lorentz tranformation that preserves three (or more) independent vectors is the identity. The only case where there are fewer than three independent $4$-vectors occurs when $N=3, n_\textrm{A}=0$, as was noted earlier as an example of an NCG. For $N>3$, there are no NCGs in the Lorentz subgroup. There is just one in the scaling subgroup, a constant phase factor \cite{copolar-invariants}.

{\it Comparison with earlier work.---}
We now return to the formulation of invariants in terms of closure traces introduced in \cite{Broderick+2020}, and 
relate it to the framework of Lorentz invariants constructed from 4-vectors developed in this paper. An elementary calculation shows that
\begin{align}
  \widehat{\bm{\mathcal{A}}} &= \frac{z^{0*} \boldsymbol{\sigma}_0 - z^{1*} \boldsymbol{\sigma}_1 - z^{2*} \boldsymbol{\sigma}_2 - z^{3*} \boldsymbol{\sigma}_3}{(\bm{\mathit{z}}\cdot\bm{\mathit{z}})^*} \, . \label{eqn:hat-advariant-4-vector}
\end{align}
Given two advariants, $\bm{\mathcal{A}}$ and $\bm{\mathcal{A}}^\prime$, we form a covariant, $\bm{\mathcal{A}}\widehat{\bm{\mathcal{A}}^\prime}$. Taking the trace and using Eq.~(\ref{eqn:hat-advariant-4-vector}), we find an invariant,
\begin{align}
  \frac{1}{2}\,\textrm{tr}\left(\bm{\mathcal{A}}\widehat{\bm{\mathcal{A}}^\prime}\right) &=\frac{\bm{\mathit{z}}\cdot \bm{\mathit{z}}^{\prime *}}{(\bm{\mathit{z}}^\prime\cdot \bm{\mathit{z}}^\prime)^*} \, , \label{eqn:Lorentz-invariant}
\end{align}
which is a ratio of Minkowskian inner products. This equation provides
the link between the closure trace formulation \cite{Broderick+2020} and the Minkowskian one presented here. Appendix~\ref{sec:short-spacing-autocorr} discusses an interesting point that the use of closely spaced elements not resolving the object, introduced in \cite{Broderick+2020} as a diagnostic, can be effectively extended to provide auto-correlation information ($n_A=1$).

We have performed numerical tests to check the analytical theory described in this paper. The expected number of independent invariants was confirmed by computing the rank of the Jacobian of the numerical partial derivatives 
of the invariants with respect to the components of the correlation matrices. We used multiple realizations of the correlations, and examined the rank of the Jacobian using its singular values. The rank agreed with our analytic results in all cases, including for $N>4$.

In the case of $N=4, n_\textrm{A}=1$, a set of covariants is exhibited in \cite{Broderick+2020} (section~3.2 and appendix~D). Because they did not account for the NCGs, they expected to find only 20 real independent invariants from 10 complex closure traces. We have numerically verified that only 18 of their 20 real invariants are independent and further that the complete and independent set consists of 21 real invariants when the NCGs are properly accounted for. This illustrates the difficulty of ensuring independence and completeness within the closure trace formalism of \cite{Broderick+2020}(see appendix~\ref{sec:closure-traces}). We conclude that the new ideas and  discussions in \cite{Broderick+2020} are valid and valuable, but the details regarding the independent loops need re-examination.

{\it Discussion.---}
The 4-vector formalism gives us an elegant criterion for deciding if the object has any polarization structure at all. In the absence of polarization structure, the 4-vectors corresponding to the true correlations have only a $0$-th component (which depends only on the Stokes total intensity components of the correlations), which renders them all collinear. As a result, $y_m^\mu$ are all collinear, since this property is preserved by Lorentz transformations and scaling. The dimension of the space spanned by the four vectors thus gives a strong statistical test for evidence of polarized structure.

This work opens up immediate applications, and areas for further investigation. In cutting-edge VLBI work on polarized emission, the availability of a full set of calibration-independent constraints will provide a valuable confirmation of derived images. Interferometry has been used, even without imaging, to study the statistics of random fields. These studies will also gain by a knowledge of closure invariants. 

In this context, the 4-vectors introduced in this paper have the advantage that they belong to a linear space, so that the same information can be encoded in linear combinations, ranked by signal to noise, via singular value decomposition. Given that direct geometric interpretation of invariants is still being explored even in the case of co-polar measurements \cite{Thyagarajan+2020c}, there is clearly much work to be done on the 4-vector approach. Finally, we anticipate that the general concept of gauge invariance we have introduced may have wider applicability, in other fields where general linear transformations of multi-channel data by unknown or ill-constrained factors have corrupted the correlations. 

\textit{Acknowledgments}.---
We thank Dom Pesce for a correspondence
clarifying the work of \cite{Broderick+2020}, Supurna Sinha, Ron Ekers, and the anonymous reviewers for a careful reading of our manuscript,
and Roshni Rebecca Samuel for her rendering of Fig.~\ref{fig:lattice}.  J.~S. acknowledges support by a grant from the Simons Foundation (677895, R.G.). 

% \bibliography{closureInvariants}% Produces the bibliography via BibTeX.

\section{The ``Hat'' operator}\label{sec:hatting} 

Some of the basic properties of the \textit{hat} operator are:
\begin{itemize}
\item The order of the inverse and conjugate-transpose operations is irrelevant,
$\widehat{\mathbf{P}} = (\mathbf{P}^{-1})^\dagger=(\mathbf{P}^\dagger)^{-1}$
\item If $\mathbf{Q}=\mathbf{P}_1\ldots\mathbf{P}_j\ldots\mathbf{P}_n$, is a  product of invertible matrices, $\mathbf{P}_j$, then
$\widehat{\mathbf{Q}}= \widehat{\mathbf{P}}_1\ldots\widehat{\mathbf{P}}_j\ldots \widehat{\mathbf{P}}_n$ retains the same order of multiplication of the `hatted' matrices, $\widehat{\mathbf{P}}_j$. 
\item $\widehat{\mathbf{C}}_{ab} \mathbf{C}_{ba} = \mathbf{I}$ (the identity matrix)
\end{itemize}

\section{Invariants with $N=3$}\label{sec:N=3}

As mentioned in the main text, the dimension of the stabilizer for any $N\ge 4$ is $\textrm{dim}(\mathcal{H})=1$. The stabilizer subgroup, $\mathcal{H}$, consists of a constant phase shift applied to the signals from all the interferometer elements, which would change $\mathbf{G}_a$ but not $\mathbf{C}_{ab}$ because correlations in standard astronomical interferometry applications depend only on the relative and not absolute phases. This means that the stabilizer contains at least the subgroup of multiplication by $\mathbf{G}_a=e^{i \theta}$ for all $a$, which is identified with $U(1)$, since such a phase factor is simply a $1\times 1$ unitary matrix. Thus, $\mathrm{dim}(\mathcal{H})\ge 1$.  In fact, in all the cases except the one below, $\mathrm{dim}(\mathcal{H})=1$.

$N=3$ presents an interesting degenerate case because it illustrates some general principles behind our construction. Consider the case without auto-correlations ($n_\textrm{A}=0$). There is just one triangle, $\Delta_{(012)}$, which gives us only one complex advariant (and so, only one complex 4-vector, $z^\mu_{12}$), which gives us just two real 4-vectors, $y^\mu_1$ and $y^\mu_2$. We can form three Lorentz invariants from Minkowski dot products, $\tilde{\mathcal{I}}_{k\ell}=\mathbf{y}_k\cdot \mathbf{y}_\ell=y^\mu_k \eta_{\mu\nu} y^\nu_\ell$, with $k,\ell=1,2$ and $\ell\ge k$, which will give us {\it two} closure invariants, $\tilde{\mathcal{I}}_{kl}/\tilde{\mathcal{I}}_{11}$, after eliminating the unknown scale factor. This is one more than that given by the formula $4N^2-12N+4n_A+1$ that works for $N>3$.

This increase in closure invariants occurs because the stabilizer now has $\textrm{dim}(\mathcal{H})=2$. To show this, we use the fact that there  is an extra one-parameter subgroup of Lorentz transformations in the plane orthogonal to the two 4-vectors, $\mathbf{y}_1$ and $\mathbf{y}_2$, which acts trivially on them (i.e., leave them unaffected). 

Corresponding to these  Lorentz  transformations, we have a one-parameter group of gauge transformations at the base $0$, $\mathbf{G}_0(s)=e^{s \mathbf{K}}$, which leaves $\bm{\mathcal A}_{12}$ unchanged 
\begin{align}\label{eqn:fixa12}
\mathbf{G}_0(s) \bm{\mathcal{A}}_{12}\mathbf{G}_0^\dagger(s) &= \bm{\mathcal{A}}_{12} \, .
\end{align}
Note that the Lorentz transformations, $\mathbf{G}_0(s)$, belong to the subgroup $\textrm{SL}(2,\mathbb{C})$ and have determinant unity, and so do not affect the scale of the vectors. The question now is whether we can find a one-parameter subgroup of gauge transformations, acting at all elements, $\{\mathbf{G}_0(s),\mathbf{G}_1(s),\mathbf{G}_2(s)\}$, such that none of the measured correlations is affected:
\begin{align}
    \mathbf{C}_{01} &= \mathbf{G}_0(s)\mathbf{C}_{01} \mathbf{G}^\dagger_1(s) \label{eqn:01} \\
    \mathbf{C}_{02} &= \mathbf{G}_0(s)\mathbf{C}_{02} \mathbf{G}^\dagger_2(s) \label{eqn:02} \\
    \mathbf{C}_{12} &= \mathbf{G}_1(s)\mathbf{C}_{12} \mathbf{G}^\dagger_2(s) \, . \label{eqn:12}
\end{align}
The answer is yes: we solve Eq.~(\ref{eqn:01}) for $\mathbf{G}_1(s)$ and Eq.~(\ref{eqn:02}) for $\mathbf{G}_2(s)$ to express them in terms of  $\mathbf{G}_0(s)$ and the measured correlations. This yields $\mathbf{G}_1(s)=\mathbf{C}_{01}^\dagger \widehat{\mathbf{G}}_0(s)\widehat{\mathbf{C}}_{01}$, and $\mathbf{G}_2(s)=\mathbf{C}_{02}^\dagger \widehat{\mathbf{G}}_0(s)\widehat{\mathbf{C}}_{02}$. It then follows from Eq.~(\ref{eqn:fixa12}) that $\mathbf{C}_{12}$ is also unchanged by the the one-parameter group of gauge transformations, $\{\mathbf{G}_0(s),\mathbf{G}_1(s),\mathbf{G}_2(s)\}$. Thus, this subgroup fixes all the measured correlations $\mathbf{C}_{ab}$, and so belongs to the stabilizer subgroup. 

This increase in the dimension of the stabilizer increases the number of invariants by one. The two invariants can also be exhibited by taking the trace of the covariant, $\bm{\mathcal{A}}_{12}\widehat{\bm{\mathcal{A}}}_{12}$, and its inverse. This is formed from the six-edged loop in which the triangle, $\Delta_1\equiv\Delta_{(012)}$, is traversed twice. While the traces of this matrix, and its inverse, are both complex, they are conjugates of each other. So, we get two real invariants from this combination. This is also clear from the representation of the trace in terms of inner products of 4-vectors, 
\begin{align}
    \mathcal{T}_{\Delta_1\Delta_1} = \frac{1}{2}\,\textrm{tr}[\bm{\mathcal{A}}_{12}\widehat{\bm{\mathcal{A}}}_{12}] &= \frac{\mathbf{z}_{12}\cdot\mathbf{z}_{12}^*}{(\mathbf{z}_{12}\cdot\mathbf{z}_{12})^*} \,.
\end{align}

This exceptional behavior disappears if an auto-correlation, $\bm{\mathcal{A}}_0$, is measured. This yields one more real 4-vector, $y^\mu_0$, so that we now have three real 4-vectors $\{y^\mu_0, y^\mu_1, y^\mu_2\}$, thereby removing the residual Lorentz gauge freedom (present earlier in the plane orthogonal to the plane containing the first two real 4-vectors, $\{y^\mu_0, y^\mu_1\}$). The number of Lorentz invariants, $\tilde{\mathcal{I}}_{k\ell}$ with $k,\ell=0,1,2$ and $\ell\ge k$, is six. After eliminating the unknown scaling factor by dividing by one of them, the number of closure invariants is $N_\mathcal{I}=5$. In matrix language, $\mathbf{G}_0(s)$ was chosen to leave the advariant $\bm{\mathcal A}_{12}$ unchanged. So if it left $\bm{\mathcal{A}}_0$ in the case $N=3$, or indeed, for $N>3$, any other advariant unchanged, it could only be by accident, since these are independent measurements. In the most general case, $\mathbf{G}_0(s)$ will no longer be a  stabilizer when $n_A=1$ and $N=3$, or when $N>3$.     
\section{Auto-correlations as the coincidence limit of Cross-correlations}\label{sec:short-spacing-autocorr}

As we have emphasised in the text, auto-correlations tend to be severely affected by additional systematic effects (which cross-correlations eliminate) and are therefore regarded as less reliable. However, when the angular size of the object is small compared to the angular resolution determined by the element spacing, we can find a way around this problem through the use of cross-correlation following the discussion in Paper~I \cite{copolar-invariants}. 

This requires an extra element $0^\prime$ in close proximity with the base element $0$. This can be provided by adding a new element, $0^\prime$, or by splitting a dense array of phased elements, as in the case of the Atacama Large Millimeter/submillimeter Array (ALMA) in the EHT/VLBI observations of M87 and Centaurus~A \cite{eht19-2,Janssen+2021}, into two independent portions that are geographically coincident and phasing them separately to act as independent elements 0 and $0^\prime$. This provides us with an extra short-spacing pair, $(0, 0^\prime)$, with separation, $D_{00^\prime}$, that is too short to resolve the object ($\theta_\textrm{obj}\ll\lambda/D_{00^\prime}$), where, $\theta_\textrm{obj}$ is the angular extent of the object, and $\lambda$ is the wavelength of observation), but distant enough that the gains, $\mathbf{G}_0$ and $\mathbf{G}_{0^\prime}$, are independent. Note that because $\mathbf{S}_{a0^\prime}\approx \mathbf{S}_{a0}$, no new true cross-correlations are provided by this additional element beyond that provided by element 0, excepting $\mathbf{S}_{00^\prime}$, where, $\mathbf{S}_{00^\prime}\approx \mathbf{S}_{00}$. Using the extra short-spacing cross-correlation measured, $\mathbf{C}_{00^\prime}$, we can construct an advariant that closely approximates that from an auto-correlation as described below.

Let us construct the advariant based on a ``thin'' triangle $\Delta_{(00^\prime a)}$, where $a\notin \{0, 0^\prime\}$: $\mathbf{C}_{00^\prime} \widehat{\mathbf{C}}_{0^\prime a} \mathbf{C}_{a0}=\mathbf{G}_0 \mathbf{S}_{00^\prime} \widehat{\mathbf{S}}_{0^\prime a} \mathbf{S}_{a0} \mathbf{G}_0^\dagger$. Under our assumption of closeness of elements 0 and $0^\prime$, this  can be written as $\approx\mathbf{G}_0 \mathbf{S}_{00} \widehat{\mathbf{S}}_{0a} \mathbf{S}_{a0}\mathbf{G}_0^\dagger=\mathbf{G}_0 \mathbf{S}_{00} \mathbf{G}_0^\dagger$ since $\widehat{\mathbf{S}}_{0a} \mathbf{S}_{a0}=\mathbf{I}$. This is effectively the same advariant that we would obtain from a direct measurement of auto-correlation at element 0. This means the earlier discussion with $n_A=1$ is applicable  and we gain four invariants. The new advariant defines a 4-vector and the four components of this 4-vector in an already established frame give us four more real invariants. The use of extra short baselines was proposed in \cite{Broderick+2020} as an error diagnostic. We note here that they can also be used to incorporate auto-correlations into the construction of invariants. 
 
\section{Connection to Closure Traces: completeness and independence}\label{sec:closure-traces}
 
In the main paper, we chose to express the closure invariants as Minkowskian inner products. This has the advantage that their independence and completeness is manifest. However, for comparison with \cite{Broderick+2020}, we express the same invariants in the language of closure traces. Let us single out two triangles (labelled $\Delta_1\equiv\Delta_{(012)}$ and $\Delta_2\equiv\Delta_{(023)}$, and their reversed forms $\nabla_1\equiv\Delta_{(021)}$ and $\nabla_2\equiv\Delta_{(032)}$) and construct the complex closure traces
\begin{align}
 \mathcal{T}_{\Delta_1\Delta_2} &= \frac{1}{2}\,\textrm{tr}[\bm{\mathcal{A}}_{12}\widehat{\bm{\mathcal{A}}}_{23}] =\frac{\mathbf{z}_{12}\cdot\mathbf{z}_{23}^*}{(\mathbf{z}_{23}\cdot\mathbf{z}_{23})^*}  \,, \nonumber\\
 \mathcal{T}_{\nabla_1\Delta_2} &= \frac{1}{2}\,\textrm{tr}[\bm{\mathcal{A}}_{12}^\dagger\widehat{\bm{\mathcal{A}}}_{23}] =\frac{\mathbf{z}_{12}^*\cdot\mathbf{z}_{23}^*}{(\mathbf{z}_{23}\cdot\mathbf{z}_{23})^*} \,, \nonumber\\
 \mathcal{T}_{\Delta_2\Delta_1} &= \frac{1}{2}\,\textrm{tr}[\bm{\mathcal{A}}_{23}\widehat{\bm{\mathcal{A}}}_{12}]   =\frac{\mathbf{z}_{23}\cdot\mathbf{z}_{12}^*}{(\mathbf{z}_{12}\cdot\mathbf{z}_{12})^*} \,, \nonumber\\
 \mathcal{T}_{\nabla_2\Delta_1} &= \frac{1}{2}\,\textrm{tr}[\bm{\mathcal{A}}_{23}^\dagger\widehat{\bm{\mathcal{A}}}_{12}] =\frac{\mathbf{z}_{23}^*\cdot\mathbf{z}_{12}^*}{(\mathbf{z}_{12}\cdot\mathbf{z}_{12})^*} \,, \nonumber\\
 \mathcal{T}_{\Delta_1\Delta_1} &= \frac{1}{2}\,\textrm{tr}[\bm{\mathcal{A}}_{12}\widehat{\bm{\mathcal{A}}}_{12}]   =\frac{\mathbf{z}_{12}\cdot\mathbf{z}_{12}^*}{(\mathbf{z}_{12}\cdot\mathbf{z}_{12})^*} \,, \nonumber\\
 \mathcal{T}_{\Delta_2\Delta_2} &= \frac{1}{2}\,\textrm{tr}[\bm{\mathcal{A}}_{23}\widehat{\bm{\mathcal{A}}}_{23}]   =\frac{\mathbf{z}_{23}\cdot\mathbf{z}_{23}^*}{(\mathbf{z}_{23}\cdot\mathbf{z}_{23})^*} \,.
\end{align}
The other pairwise advariant combinations not written here can be shown to be dependent on the above.

It would appear that we have twelve real invariants from these six complex closure traces. However, they have further dependencies satisfying the following three real relations:
\begin{align}
    \left|\frac{\mathcal{T}_{\nabla_1\Delta_2} \mathcal{T}_{\Delta_2\Delta_1}}{\mathcal{T}_{\Delta_1\Delta_2}\mathcal{T}_{\nabla_2\Delta_1}}\right| &= 1 \,, \label{eqn:relation1} \\
    \frac{\mathcal{T}_{\Delta_1\Delta_2}^* \mathcal{T}_{\nabla_2\Delta_1}\mathcal{T}_{\Delta_2\Delta_2}}{\mathcal{T}_{\nabla_1\Delta_2}\mathcal{T}_{\Delta_2\Delta_1}\mathcal{T}_{\Delta_2\Delta_2}^*} &= 1 \,,     \label{eqn:relation2} \\
    \textrm{and}\quad \frac{\mathcal{T}_{\nabla_1\Delta_2} \mathcal{T}_{\Delta_2\Delta_1}^*\mathcal{T}_{\Delta_1\Delta_1}}{\mathcal{T}_{\Delta_1\Delta_2}\mathcal{T}_{\nabla_2\Delta_1}\mathcal{T}_{\Delta_1\Delta_1}^*} &= 1 \,.     \label{eqn:relation3}
\end{align}
This reduces the number of real invariants to 9, made from the advariants $\bm{\mathcal{A}}_{12}$ and $\bm{\mathcal{A}}_{23}$ alone. In addition, each of the remaining triangles, $\Delta_k$ with $k\ge 3$, contributes the following four complex invariants constructed by taking the traces of the products of each $\bm{\mathcal{A}}_{\Delta_k}$ with $\bm{\mathcal{A}}_{\Delta_1}$ and  $\bm{\mathcal{A}}_{\Delta_2}$, and their Hermitian adjoints, 
\begin{align}
\mathcal{T}_{\Delta_1\Delta_k} = \frac{1}{2}\,\textrm{tr}[\bm{\mathcal{A}}_{\Delta_1}\widehat{\bm{\mathcal{A}}}_{\Delta_k}] &\,,\, \mathcal{T}_{\Delta_2\Delta_k} = \frac{1}{2}\,\textrm{tr}[\bm{\mathcal{A}}_{\Delta_2}\widehat{\bm{\mathcal{A}}}_{\Delta_k}] \,, \nonumber\\
\mathcal{T}_{\nabla_1\Delta_k} = \frac{1}{2}\,\textrm{tr}[\bm{\mathcal{A}}_{\Delta_1}^\dagger\widehat{\bm{\mathcal{A}}}_{\Delta_k}] &\,,\, \mathcal{T}_{\nabla_2\Delta_k} = \frac{1}{2}\,\textrm{tr}[\bm{\mathcal{A}}_{\Delta_2}^\dagger\widehat{\bm{\mathcal{A}}}_{\Delta_k}] \,.
\end{align}
As earlier, we can express these in terms of ratios of Lorentz-invariant Minkowskian inner products, and in this case it is easy to see that the four complex invariants are indeed independent. This gives us eight real invariants per additional triangle, $\Delta_k$, for $3\le k\le (N-1)(N-2)/2$, in full agreement with the counting based purely on Lorentz invariants. Also, note that the argument in the main text based on Lorentz invariants is considerably simpler than the one described above. 

\section{Holonomy and the Choice of Base Vertex}\label{sec:holonomy}

In gauge theory, it is usual to talk about the holonomy group  at a point, $P$, which gives us the parallel transport along loops $\Gamma$ pinned at $P$. One can multiply two loops by traversing them in succession and the holonomy of the product of the loops is the product of the holonomy of the individual  loops. In unitary lattice gauge theories (and more generally gauge theory on a graph), there are elementary plaquettes and all loops can be broken into these elementary plaquettes, which are both independent and complete. In our problem, the gauge group was not unitary, and the even holonomies were quadrilateral and therefore not independent. We got around this difficulty by using triangular advariants which are both independent and complete. The advariants are not themselves holonomies, but can be used to construct all holonomies. 

The holonomy group depends on the base vertex, but only changes by conjugation. So, the choice of the base vertex is not important from the mathematical point of view, and we have arbitrarily singled out one array element (indexed 0) as the base vertex. In practice, the choice of the base vertex may be guided by other  considerations like the quality of data received at different elements. A final remark about nomenclature: the word ``holonomy'' which is used in the mathematical literature is known to physicists as a ``Wilson loop''. In the geometric phase literature, it is sometimes referred to as an ``{\it anholonomy}'', following Michael Berry's just observation \cite{Berry1987a,Berry1987b} that we are really describing a failure of integrability and therefore a lack of ``wholeness''. This usage is consistent with ``holonomic and anholonomic constraints'' of classical mechanics. 

\end{document}